\documentclass[conference]{IEEEtran}
\IEEEoverridecommandlockouts
\usepackage{cite}
\usepackage{amsmath,amssymb,amsfonts}
\usepackage{algorithmic}
\usepackage{graphicx}
\usepackage{textcomp}
\usepackage{xcolor}

\usepackage{subfig}

\def\BibTeX{{\rm B\kern-.05em{\sc i\kern-.025em b}\kern-.08em
    T\kern-.1667em\lower.7ex\hbox{E}\kern-.125emX}}
\begin{document}
\title{Real-time FPGA Implementation of CNN-based Distributed Fiber Optic Vibration Event Recognition Method\\ 
}
\author{
    \IEEEauthorblockN{Zhongyao Luo, Zhao Ge, Hao Wu, Ming Tang}
    \IEEEauthorblockA{\textit{ National Laboratory for Optoelectronics (WNLO) \&}}
    \IEEEauthorblockA{\textit{National Engineering Laboratory for Next Generation Internet Access System}}
    \IEEEauthorblockA{\textit{School of Optical and Electronic Information}}
    \IEEEauthorblockA{\textit{Huazhong University of Science and Technology}}
    \IEEEauthorblockA{\textit{Wuhan, China}}
    \IEEEauthorblockA{zluo@hust.edu.cn, d202280977@hust.edu.cn, wuhaoboom@hust.edu.cn, tangming@mail.hust.edu.cn}
}

\maketitle

\begin{abstract}

Utilizing optical fibers to detect and pinpoint vibrations, Distributed Optical Fiber Vibration Sensing (DVS) technology provides real-time monitoring and surveillance of wide-reaching areas.
This field has been leveraging Convolutional Neural Networks (CNN). Recently, a study has accomplished end-to-end vibration event recognition, enabling utilization of CNN-based DVS algorithms as real-time embedded system for edge computing in practical application situations. 
Considering the power consumption of central processing unit (CPU) and graphics processing unit (GPU), and the inflexibility of application-specific integrated circuit (ASIC), field-Programmable gate array (FPGA) is the optimal computing platform for the system. 
This paper proposes to compress pre-trained network and adopt a novel hardware structure, to design a fully on-chip, pipelined inference accelerator for CNN-based DVS algorithm, without fine tuning or re-training.
This design allows for real-time processing with low power consumption and system requirement.
An examination has been executed on an existing DVS algorithm based on a 40-layer CNN model comprising 2.7 million parameters. It is completely implemented on-chip, pipelined, with no reduction in accuracy.

\end{abstract}


\section{Introduction}


Distributed optical fiber vibration sensing technology (DVS) enable continuous monitor for vibration events along the fiber at the lengths of dozens of kilometers, which has been applied in the various fields including geophysical\cite{Miah2017ARO}.
Since 2018, convolutional neural network (CNN) has been employed for DVS event recognition\cite{10.1117/1.OE.57.1.016103}. 
The latest research has revealed the possibility for CNN-Based DVS algorithm to achieve end-to-end interference vibration recognition, which is to process and analyze sensing data collected from optical fiber directly, without need for buffering all sampled data\cite{Ge2022HighAccuracyEC}.

\begin{figure}[ht]
\centerline{
\includegraphics[width = 0.3\textwidth]{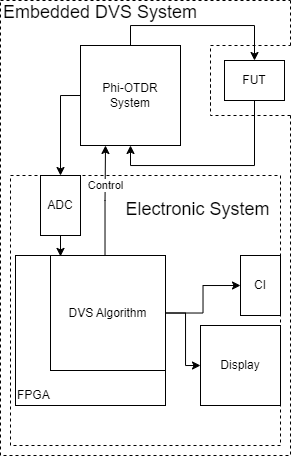}
}
\caption{Schematic diagram of DVS embedded system. CI: communication interface, used to communicate with other devices. ADC: analog-to-digital converter. FUT: fiber under test.}
\label{embed}
\end{figure}

Based on previous work on embedded  optical time-domain reflectometry (OTDR) system\cite{6963869}, it shows that it is feasible to assemble a total system that can effectively detect vibration events in real-time. Taking into account real-world situations, such as geophysical, the ideal system would be an embedded system, allowing for edge computing. It brings the demand for energy efficiency and costs, which makes graphics processing unit (GPU), which have high prices and energy consumption, unsuitable for the inference\cite{9609699}, a bad choice.  
In light of the quickly changing CNN algorithms, application-specific integrated circuit (ASIC) like Eyeriss\cite{7738524}, with their constrained customizability, are not an ideal choice. Field-Programmable gate array (FPGA), with a low cost and a low power consumption in comparison to GPUs, and a high flexibility in comparison to ASICs, become the primary option. Additionally, algorithm implementation utilizing pipeline architecture on FPGA allows for real-time processing. The expected embedded system is illustrated in figure\ref{embed}.

To this point, the remaining task is the FPGA implementation of CNN, which is also a popular research topic in recent years.
Conventional FPGA-based CNN accelerator typically involves dynamic random-access memory (DRAM) for storage, introducing high energy consumption\cite{9556422}. With the goal of further decreasing energy consumption and system requirement, it is obvious that the need for DRAM should be eliminated, that is, the network should be implemented using solely on-chip memory. To accomplish this, the idea of two well-known techniques, pruning and quantization, can be utilized\cite{8114708}.

Drawing on the concept of ShiftAddNet\cite{NEURIPS2020_1cf44d79}, this paper combines pruning and quantization to suggest replacing convolution layers in the network with a combination of shift and add layers, and substituting weights with the numbers of bits to be shifted in each shift layer, with the intention of minimizing the size of the network and reducing its hardware requirements. The proposed scheme is intended to be carried out on a pre-training model of CNN, without fine-tuning of the network structure, and without a retraining program, which clearly requires less time and work than conventional solutions\cite{9556422}. It has been evaluated against an existing CNN-based DVS algorithm. The pre-trained CNN model and original dataset are provided by the author\cite{Ge2022HighAccuracyEC}.

\section{Implementation Scheme} 

\subsection{Hardware Structure}
\paragraph{Basic Convolution Layer Structure}  The operation of a neuron in a conventional convolution layer can be expressed as below, with \textit{o} being the output, \textit{i} being the elements in the input matrix, and \textit{w} being the elements in the weight matrix.

\begin{equation}
\label{eq_1}
\textit{o} = \sum \textit{i} \cdot \textit{w}
\end{equation}

The operation can be seen as a combination of element-wise multiplication between the input data matrix and weight matrix, and subsequent summation of all the products. It is evident that the arithmetic operations mainly involved here are multiplication and addition.

\begin{figure}[ht]
\centerline{
\includegraphics[width = 0.45\textwidth]{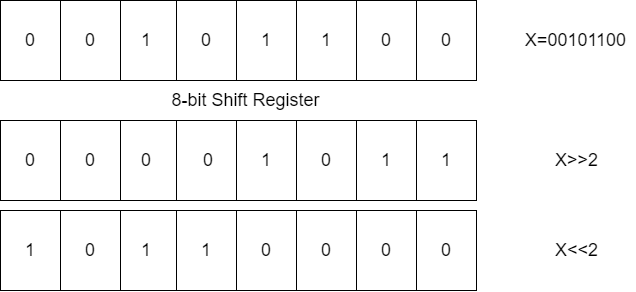}
}
\caption{Example of shift operation.}
\label{shift}
\end{figure}

\paragraph{Shift-Add Structure} 
 A signed binary number can be understood as the addition of multiple numbers which are powers of two. It is then multiplied by $s \in \{-1, 1\}$ to represent the sign, as shown in equation below. $2^{p}$ can be seen as bit-wise shift operations, with the integer p representing the number of shifts and shift direction. The shift direction is either right, when $p < 0$, or left, when $p > 0$.  Figure \ref{shift} gives an example of shift operation.

\begin{equation}
\label{eq_2}
    w = s \cdot \sum 2^{p}
\end{equation}

The convolution equation can be expressed as follows. It shows that the neuronal operation can be divided into series of shifting and addition operations.

\begin{equation}
\label{eq_3}
o = \sum s \cdot i \cdot \sum 2^{p} = \sum s \cdot \sum i \cdot 2^{p}
\end{equation}

\begin{figure*}[ht]
\centerline{
\includegraphics[width = \textwidth]{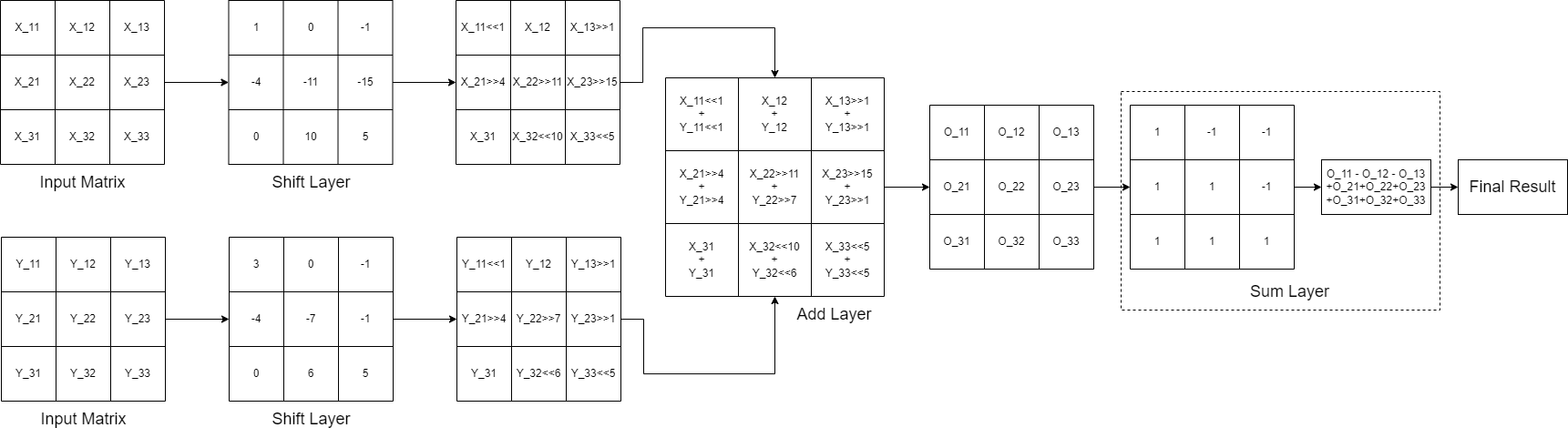}
}
\caption{Example of Shift-Add structure with two shift layers. Parameters in shift layer are randomly generated used for demonstration. X,Y: input matrix. O: output data matrix from add layer.}
\label{SA}
\end{figure*}

In contrast to the shift and add layer approach suggested in ShiftAddNet\cite{NEURIPS2020_1cf44d79}, the design proposed in this paper employs multiple shift layers. Element-wise addition layers are presented to carry out element-wise addition between shift layer outputs. All the products are summed together in the global add layer, based on the sign data. These layers are arranged in the same way as in the equation \ref{eq_3}. The proposed design should contain N shift layers accompanied by N-1 element-wise accumulation layers and one global add layer. (For brevity, only the quantity of shift layers will be referred in the rest parts of the paper.) An example of the proposed Shift-Add structure with two shift layers is illustrated in figure \ref{SA} It is evident that, with an adequate number of shift layers, the recommended structure can attain the same expressive capability as a standard convolution layer. 

From hardware perspective, the operations involved are shifting and addition. Shift operation offers faster computation speed and lower energy expenditure when compared to multiplication\cite{9522876}. Additionally, the parallelism of traditional FPGA-based CNN accelerators is often constrained by the number of multipliers which necessitates the utilization of DSP resources. Thus, the computations of layers are typically performed in time-multiplexed ways, which limits the overall throughput of the accelerator\cite{9556422}.

\begin{figure}[htbp]
\centering
\subfloat[Multiply and Accumulation Unit(MAC)]{\includegraphics[width = 0.25\textwidth]{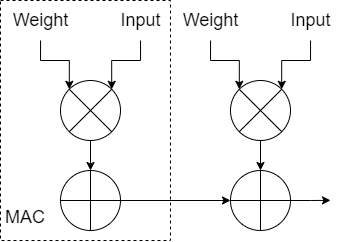}} 
\subfloat[3-MAC-based CNN accelerator]{\includegraphics[width = 0.25\textwidth]{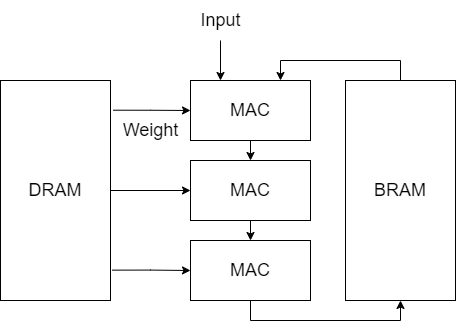}}
\caption{Simple FPGA-based CNN accelerator structure}
\label{MAC}
\end{figure}

A simple FPGA-based CNN accelerator is shown in figure \ref{MAC}. If a one-dimensional convolution layer with 3 weights is implemented using the accelerator, the resulting implementation will be fully-parallel, with each weight assigned to an individual MAC.
The input data will be sequentially processed in each MAC to obtain the final result. It means that the implementation is pipelined, and able to continuously process the input data, while keeping producing corresponding results, achieving real-time computation naturally.
If the convolution layer has more weights, the computation process will be time-multiplexed.  The input data is first computed with MACs using first 3 weights. The result is stored in block random-access memory (BRAM) for next round of computation with next set of 3 weights read from DRAM. The process repeats until all weights have been iterated through. To meet the requirement for pipelining, it brings strict timing constraints for each operation, so that the total computation time is less then the time interval between input data.

The presented design involves no DSP slices, which means that the number of Shift-Add unit implemented on chip is not limited by DSP resources, bringing higher level of parallelism, enabling fully parallel implementation for network with more weights. Furthermore, with higher computation speed, the network implemented using Shift-Add structure can meet the timing requirement for pipelining more easily. Accordingly, it is feasible to integrate the entire network onto the chip, using proposed design, in a pipelined manner, allowing for real-time processing of input data.

\subsection{Network Compression}

Pruning is done on the basis that the network is generally over-parameterized, which indicates that there are redundancies in the parameters of the network, which are able to be eliminated. This procedure of discarding redundant information is known as pruning. Quantization involves transforming data to a restricted range which can be represented with fewer bits, while keeping a minimum quantization error \cite{8114708}. Data is typically mapped with 8 bits of integer, because the underlying operations in the integer range are much simpler than the operations in the floating-point range, thereby reducing the need for resources and speeding up computation\cite{8578384}.

If the aforementioned structure is adopted to substitute convolution layers, the weights should be replaced with multiple integers, as the \textit{p} in equation \ref{eq_2}, so that they can be easily used by the shift layers.

A simple encoding system is used here, to simplify the decoding process, where the parameters are represented by numbers, which correspond to the relative position of zeros or ones, depending on the majority digit of the number. For example, the unsigned binary number 1110101 can be expressed as two integers, 2 and 4, in addition to the details of the majority digit in the figure and its width.

utilizing the encoding process, the parameter is mapped into multiple integers in relatively small data range with lower requirement for number of bits used for representation. For example,  a 32-bit floating-point figure can be characterized with multiple five-bit integers. However, in the worst case, it may demand 16 five-bit variables to characterize each parameter, with the requirement for 16 shift layers to substitute one original convolution layer. It is evident that encoding can fulfill the role of a compression scheme for parameters, provided that the number of variables used is restricted. The limitation in this example would be $\lfloor 32 \div 6 \rfloor = 5$.

Here, the concept of pruning is adopted, not towards the nodes in the network structure, but the number of variables used to represent parameters in the convolutional layers. An efficient approach is to evaluate all possible permutations of the number of variables involved in the different layers on the test dataset, in order to identify the combination with the best performance. 
Since pruning is considered to be a process removing redundant information from the model, it is better to first use the same number of variables to represent parameters in different convolution layers, then to change the number and to evaluate the models. It is advised to select the most suitable model based on the number of shift layers required and its performance.
Based on the selected model, new models are derived through changing the number of variables used in different convolution layers, and then evaluated, searching for a better model. In order to make the most efficient use of our time and efforts, it is suggested to cease the process when a suitable model is attained.

\section{Results and Analysis}

\begin{figure}[ht]
\centerline{
\includegraphics[width = 0.45\textwidth]{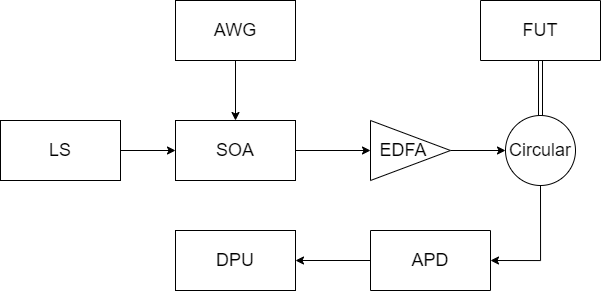}
}
\caption{$\varphi$-OTDR system structure. LS: light source, SOA: semiconductor optical amplifier, AWG: arbitrary waveform generator, EDFA: erbium-doped fiber amplifier, FUT: fiber under test, APD: avalanche photodetector, DPU: data processing unit}
\label{phi}
\end{figure}

The scheme is evaluated on a CNN-based DVS algorithm designed for recognition of vibration events, including ones caused by excavator, hammer, and air pick\cite{Ge2022HighAccuracyEC}. The algorithm is based on a direct-detection phase-sensitive OTDR ($\varphi$-OTDR), as shown in figure \ref{phi}. The provided datasets are sampled on a 12.5-meter-long fiber at 80 MSa/s, in a system with a spatial resolution of 5 meters and a pulse repetition rate of 1kHz. The algorithm first accumulates and averages over every four samples. Then it takes the resulting data in 256 ms as the input for the CNN used, which means that the input data matrix is at the size of $256\times11$.

Hardware design is implemented using Xilinx HLS and tested on ZCU15eg FPGA chip. The chip provides total BRAM of 26.2 Mb and maximum distributed ram of 11.3 Mb, giving 37.3 Mb maximum on-chip storage. Only FPGA part is used in this work. 

\subsection{Compression Result}
To begin with, each convolution layer is substituted with the same number of shift layers. Models of varying number of shift layers are then measured against the test dataset. The results are shown in table \ref{tab_1}. The results indicate that 6 layers are sufficient for achieving the same results as the initial model, which indicates that six variables are enough for storing all the necessary information. The models with 2, 4, and 5 shift layers provide equivalent results, implying that the two-shift-layer model holds most essential information.

Based on the selected model, new models are derived through choosing different convolution layer to be replaced by only one shift layer, and then evaluated on test dataset. The result is shown in table \ref{tab_2}. The original network has fifteen convolutional layers and one fully-connected layer. Therefore, the Layer Number in the table shows which layer is chosen. The table indicates that models with convolution layer 3, 4, or 6 substituted by a single shift layer give an enhanced performance in comparison with the original network. The model with the third convolution layer substituted by one shift layer, and the rest altered with two shift layers has been tested on the train dataset. The result as depicted in the table \ref{tab_3} demonstrates that the majority of the information eliminated during the process is redundant.

After a thorough investigation of the value range of the variables employed to characterize the parameters in the original model, 6-bit signed integer format is adopted to depict the variables. Up to this stage, the exact size of the two-shift-layer model can be precisely calculated. The total number of parameters in original model is 2777283. The parameter number of the third convolution layer is 36864. The total size of the compressed model is $2777283 \times 12 - 36864 \times 6 = 33106212 bits \approx 31.57 Mb$, less than the maximum storage size of 37.6 Mb of target FPGA. The original model size is 10.59 MB. Thus, the compression rate is 37.26\%. It is evident that the selected model meets the fundamental requirement, thus making it the final model.

\begin{table}[htbp]
\caption{Recognition Result on Test Dataset}
\begin{center}
\begin{tabular}{|c|c|c|c|c|}
\hline
Number of Shift&\multicolumn{3}{|c|}{Identification Category/\%} & Recognition \\
\cline{2-4}  
Layer used & Excavator & Hammer & Air Pick &  Rate/\% \\
\hline
Original Network& 100 & 97.76 & 99.70 & 99.20 \\ 
\hline
1 & 100 & 93.28 & 82.42 & 91.31 \\ 
\hline
2 & 100 & 97.39 & 99.70 & 99.09\\ 
\hline
3 & 100 & 95.90 & 99.70 & 98.63 \\ 
\hline
4 & 100 & 97.39 & 99.70 & 99.09 \\ %
\hline
5 & 100 & 97.39 & 99.70 & 99.09 \\
\hline
6 & 100 & 97.76 & 99.70 & 99.20 \\ 
\hline
7 & 100 & 97.76 & 99.70 & 99.20 \\
\hline
8 & 100 & 97.76 & 99.70 & 99.20 \\
\hline
\end{tabular}
\label{tab_1}
\end{center}
\end{table}

\begin{table}[htbp]
\caption{Recognition Result on Test Dataset}
\begin{center}
\begin{tabular}{|c|c|c|c|c|}
\hline
Layer Number &\multicolumn{3}{|c|}{Identification Category/\%} & Recognition \\
\cline{2-4}  
 & Excavator & Hammer & Air Pick &  Rate/\% \\
\hline
Original Network& 100 & 97.76 & 99.70 & 99.20 \\ 
\hline
1 & 100 & 98.88 & 79.90 & 88.91 \\ 
\hline
2 & 100 & 97.76 & 98.79 & 98.86 \\ 
\hline
5 & 100 & 97.76 & 99.70 & 99.20 \\
\hline
3,4,6 & 100 & 98.13 & 99.70 & 99.31 \\ 
\hline
7-8, 10-11 & 100 & 97.76 & 99.39 & 99.09 \\ 
\hline
9, 12-16 & 100 & 97.39 & 99.70 & 99.09 \\ 
\hline
\end{tabular}
\label{tab_2}
\end{center}
\end{table}

\begin{table}[htbp]
\caption{Recognition Result on Train Dataset}
\begin{center}
\begin{tabular}{|c|c|c|c|c|}
\hline
Class &\multicolumn{3}{|c|}{Identification Category/\%} & Recognition \\
\cline{2-4}  
 & Excavator & Hammer & Air Pick &  Rate/\% \\
\hline
Original Network& 100 & 99.97 & 100 & 99.99 \\ 
\hline
Compressed Network & 100 & 99.97 & 100 & 99.99 \\
\hline
\end{tabular}
\label{tab_3}
\end{center}
\end{table}

\begin{figure}[htbp]
\subfloat[Network Structure]{\includegraphics[width = 0.5\textwidth]{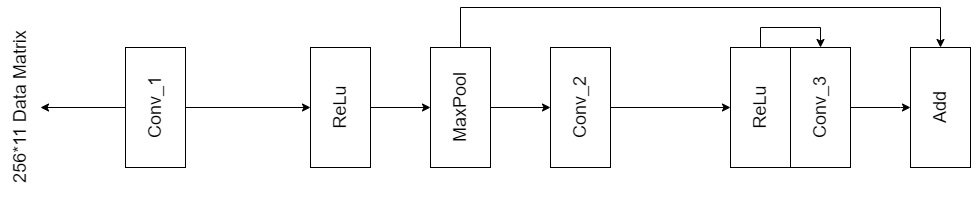}} \\
\subfloat[Hardware Structure]{\includegraphics[width = 0.5\textwidth]{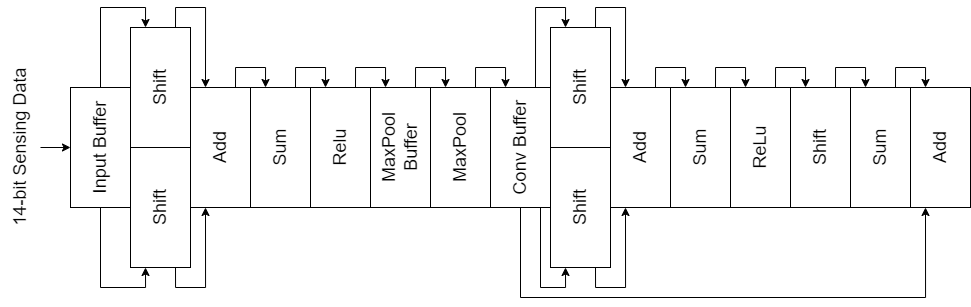}}
\caption{Structure of the design. Only the first 7 layers are shown here, since the remaining part are in similar structure.}
\label{net}
\end{figure}

\subsection{Hardware Implementation}
A library featuring templates of core layers has been created for this task. The compressed network is less than the upper boundary of on-chip storage, yet still greater than the total BRAM size. Consequently, the additional information is stored using distributed ram implemented as LUTRAM. 

The design is pipelined, continuously taking in fourteen-bit sensing data, while producing the recognition result directly. 
The diagram in figure \ref{net}, shows the hardware structure of network with reference to the original network structure. 
The implementation result of the complete design given by Vivado is shown in table \ref{tab_4}. The total number of clock cycle is 99313805, while the design is running at 250 Mhz, giving the total latency of around 397 ms. The total on-chip energy expenditure is 10.023W, which is consistent with the output power capacity of the majority of power banks available\cite{PowerBank}. Moreover, the system requirement of the chip can be diminished beyond the typical minimal system, since there is no need for DRAM\cite{MinimalSys}.

\begin{table}[htbp]
\caption{Hardware Resource Utilized}
\begin{center}
\begin{tabular}{|c|c|c|c|}
\hline
Resource & Utilization & Available & Utilization/\% \\
\hline
LUT & 155380 & 341280 & 45.5286 \\
LUTRAM & 6449 & 184320 & 3.4988065\\
FF & 25208 & 682560 & 3.6931553 \\
BRAM & 733.5 & 744 & 98.588715 \\
IO & 2 & 328 & 0.60975605 \\
BUFG & 3 & 404 & 0.7425743 \\
PLL & 1 & 8 & 12.5 \\
\hline
\end{tabular}
\label{tab_4}
\end{center}
\end{table}

After conducting an analysis of the design, the primary factor impeding performance is the buffering system. The present design adopts a straightforward buffering mechanism that caches all outcomes from the preceding layer and then injects the data into the computation, leading to a considerable increase in the total delay.

\section*{Conclusion}
This paper proposes a scheme for realizing a fully on-chip, pipelined, FPGA-based implementation of CNN-based DVS algorithms. The proposed scheme is targeted to pre-trained network, involves no fine tuning or re-training. An end-to-end CNN-based DVS algorithm is utilized to examine the scheme. The result reveals the potential of an embedded DVS system for edge computing in practical application.



\bibliographystyle{ieeetr}
\bibliography{acp_2023}

\end{document}